\documentclass[numberedappendix]{emulateapj}

\usepackage{epsfig, natbib, color}
\tightenlines

\countdef\decade=200
\advance\decade by \year
\countdef\hours=201	
\advance\hours by \time
\divide\hours by 60
\countdef\mins=202
\advance\mins by \hours
\multiply\mins by 60
\multiply\hours by 100
\countdef\miltime=203
\advance\miltime by \hours
\advance\miltime by \time
\advance\miltime by -\mins

\newcommand{\hi}{{\rm H\,\textsc{i}}}
\newcommand{\taud}{\tau_{\rm d}}
\newcommand{\sigmad}{\sigma_{\rm d}}

\defcitealias{krumholz08c}{KMT08}
\defcitealias{krumholz09a}{KMT09}

\begin{document}

\title{On the Absence of High Metallicity-High Column Density Damped Lyman $\alpha$ Systems: Molecule Formation in a Two-Phase Interstellar Medium}


\author{Mark R. Krumholz}
\affil{Department of Astronomy and Astrophysics, University of California, Santa Cruz, CA 95064, USA}
\email{krumholz@ucolick.org}

\author{Sara L. Ellison}
\affil{Department of Physics and Astronomy, University of Victoria, BC, V8P 5C2, Canada}

\author{J. Xavier Prochaska}
\affil{Department of Astronomy and Astrophysics, University of California, Santa Cruz, CA 95064, USA}

\author{Jason Tumlinson}
\affil{Space Telescope Science Institute, Baltimore, MD 21218, USA}

\begin{abstract}
We argue that the lack of observed damped Lyman $\alpha$ (DLA) systems that simultaneously have high \hi\ columns densities and high metallicities results naturally from the formation of molecules in the cold phase of a two-phase atomic medium in pressure balance. Our result applies equally well in diffuse systems where the ultraviolet radiation field is dominated by the extragalactic background and in dense star-forming ones, where the local radiation field is likely to be orders of magnitude higher. We point out that such a radiation-insensitive model is required to explain the absence of high column - high metallicity systems among DLAs observed using gamma-ray burst afterglows, since these are likely subjected to strong radiation fields created by active star formation in the GRB host galaxy. Moreover, we show that the observed relationship between the maximum atomic gas column in DLAs sets a firm upper limit on the fraction of the mass in these systems that can be in the warm, diffuse phase. Finally, we argue that our result explains the observed lack of {\it in situ} star formation in DLA systems.
\end{abstract}

\keywords{galaxies: ISM --- gamma rays: bursts --- ISM: molecules --- ISM: structure --- quasars: absorption lines --- stars: formation}

\section{Introduction}
\label{intro}

Damped Lyman $\alpha$ (DLA) systems are clouds of neutral atomic hydrogen with column densities $N(\hi) \geq 2 \times 10^{20}$ cm$^{-2}$ that are detected as absorbers against bright background quasars (QSOs) or gamma-ray bursts (GRBs) \citep{wolfe05a, prochaska05a, prochaska07a, prochaska08a}. These systems comprise the bulk of the neutral gas in the universe at redshifts up to at least $z \sim 5$. Because of their ubiquity, the study of DLAs provides vital clues to the distribution of gas and metals, and potentially also star formation, in the universe. 

Observations of DLAs show a clear zone of exclusion: none are observed with both high metallicity and high \hi\ column density. One possible explanation for this effect is that lines of sight with large column densities and metallicities produce large dust extinctions that might lead to exclusion of the background QSO from optically-selected samples \citep{boisse98a, prantzos00a}. However, statistical analysis of the optically-selected QSO-DLA sample suggests that few DLAs are missed due to extinction \citep{pontzen09a}, and radio-selected QSO-DLA samples do not differ significantly from optically-selected ones \citep{ellison01a, ellison05a, akerman05a, jorgenson06a}. 

An alternative hypothesis to explain the zone of exclusion is that above some threshold total hydrogen column density, which decreases with increasing metallicity, gas forms molecular hydrogen that is not detectable in Ly $\alpha$ absorption \citep{schaye01a, hirashita05a, hirashita06a}. Large amounts of molecular gas are not observed in these DLAs because molecular clouds have a very small covering fraction and are unlikely to be seen along random sightlines \citep{zwaan06a}, although in some cases trace amounts of molecular gas have been detected \citep[e.g.][]{ledoux03a, noterdaeme08a}. In one GRB-DLA a substantial column of molecular hydrogen has been detected \citep{prochaska09a}, and, as we show below, this system is unique among DLAs in its high column density and metallicity.

While the molecule-formation hypothesis avoids the problems of the dust bias explanation, it also has significant weaknesses. In DLAs we can observe only column density and metallicity, so previous authors have been forced to assume values, which may be incorrect, for other quantities such as total gas volume density and radiation field that influence the molecule fraction. For example, \citet{schaye01a} assumes a Lyman-Werner (LW) radiation field within a factor of 3 of the \citet{haardt01a} UV background at $z=3$, which is 50 times smaller than the Solar neighborhood value, while the weakest radiation field that \citet{hirashita06a} consider is 2 times {\it larger} than in the Solar neighborhood. Observationally-inferred radiation fields in DLAs span this full range and more \citep{tumlinson07a, wolfe08a}, so neither assumed value works for the full DLA population.

Our goal in this paper is to explain the zone of exclusion using the \citet[hereafter KMT08 and KMT09]{krumholz08c, krumholz09a} theory of the atomic to molecular transition in galaxies, which does not depend on unobservable quantities such as the gas volume density and UV radiation field. Instead, \citetalias{krumholz09a} show that to good approximation the molecular fraction in a cloud depends only on its column density and a single dimensionless parameter, which combines the volume density, radiation intensity, H$_2$ formation rate coefficient, and dust opacity, and that the two-phase nature of the atomic interstellar medium imposes strong constraints on the values this parameter can take. This model explains the observed molecular fractions and star formation rates in nearby galaxies (\citetalias{krumholz09a}; \citealt{krumholz09b}), and in \S~\ref{exclusion} we apply it to DLAs. In \S~\ref{twophase} we demonstrate that the observed zone of exclusion sets strong constraints on the fraction of warm atomic gas in DLAs. Finally, in \S~\ref{disc} we discuss the implications of our work.

\section{Molecule Formation and the Zone of Exclusion}
\label{exclusion}

\subsection{Theoretical Model of ${\rm H}_2$ Formation}
\label{theory}

We refer readers to \citetalias{krumholz08c} and \citetalias{krumholz09a} for a full derivation of the formalism, and here simply summarize. Consider a spherical cloud of gas immersed in a uniform, isotropic dissociating radiation field. The outer parts are kept predominantly atomic by the radiation, but as one moves inward dissociating photons are absorbed by H$_2$ molecules and dust grains. At some depth into the cloud all dissociating photons have been absorbed, and there is a sharp transition to predominantly molecular material. The fraction of the cloud radius at which this transition occurs depends on two dimensionless numbers: $\taud=N \sigmad$, the dust optical depth of the cloud, and $\chi = f_{\rm diss} \sigmad c E_0^* / (n_{\rm HI} \mathcal{R})$, the dimensionless strength of the radiation field. Here $N$ is the center-to-edge column density of the cloud including atomic and molecular material, $\sigmad$ is the dust cross section per H nucleus to photons in the dissociating LW bands, $f_{\rm diss}\approx 0.1$ is a quantum-mechanical constant describing the approximate probability of dissociation per absorption, $E_0^*$ is the number density of photons in the LW bands of the background dissociating radiation field, $n_{\rm HI}$ is the number density of gas in the atomic envelope of the cloud, and $\mathcal{R}$ is the rate coefficient (in units of length$^3$ time$^{-1}$) describing formation of hydrogen molecules on the surfaces of dust grains. In the limit $N\rightarrow\infty$, the \hi\ column density $N(\hi)$ approaches a finite maximum value that depends only on $\chi$ and on the dust cross section $\sigmad$ (which in turn depends on metallicity) in the atomic shielding region around the cloud. In effect, a fixed optical depth of material can absorb the entire LW photon flux, so any additional gas is molecular. This produces the observed zone of exclusion: a metallicity-dependent maximum \hi\ column.

\citetalias{krumholz09a} point out that $\chi$ cannot vary strongly between galaxies. Both $\sigmad$ and $\mathcal{R}$ measure the total surface area of dust grains in the gas, so $\sigmad/\mathcal{R}$ is insensitive to changes in dust abundance or size distribution, and thus varies little with environment. Moreover, in a two-phase atomic medium, the \hi\ density $n_{\rm HI}$ that determines when molecules form is the density of the cold neutral phase, $n_{\rm CNM}$. This is because the effective LW opacity of a fluid element provided by H$_2$ absorption is proportional to its density (cf.\ equation 8 of \citetalias{krumholz08c}). Thus the low density of the warm phase guarantees that it provides negligible self-shielding compared to the cold gas, and we care almost exclusively about $n_{\rm CNM}$. Of course the WNM does contain dust, but gas only becomes predominantly molecular once the ambient UV radiation field has been attenuated by a factor of $\sim 10^3-10^4$. This level of dust attenuation requires a color excess $E(B-V) \ga 0.6$, larger than the highest known $E(B-V)$ \citep[e.g.][]{junkkarinen04a, wild06a} and $\sim 2$ orders magnitude above the mean \citep{ellison05a, vladilo08a}. Thus we can safely neglect the contribution of WNM dust shielding in favor of CNM self-shielding.
The ratio $E_0^* / n_{\rm CNM}$ is tightly constrained by the thermodynamics of the gas and the requirement of pressure balance between the two phases \citep{wolfire03}; a reasonable approximation is $E_0'/n_{\rm CNM} \approx (1 + 3.1 Z'^{0.365})/93$ cm$^3$, where $Z'$ and $E_0'$ are the metallicity and FUV radiation intensity $E_0^*$ normalized to their values in the Solar neighborhood.\footnote{Following \citet{draine78}, we take the LW radiation intensity in the Solar neighborhood to have a value that produces a free-space dissociation rate of $5.43\times 10^{-11}$ s$^{-1}$; this is $1.6$ times the \citet{habing68} field.} Together, with the invariance of $\sigmad/\mathcal{R}$, this (weak) dependence of $E_0^*/n_{\rm CNM}$ on $Z'$ gives a dimensionless radiation intensity $\chi\approx 0.77 (1 + 3.1 Z'^{0.365})$ that depends only on the metallicity of the gas. 

The implication of this result is that, in the CNM, the dimensionless radiation strength $\chi$ does not depend on the absolute FUV radiation field. Any change in radiation intensity induces a countervailing change in density. This is why there is a zone of exclusion for all DLAs despite the huge range in radiation intensities inferred within them. To calculate this effect quantitatively, if we assume that dust opacity $\sigmad$ scales with metallicity, then in the KMT formalism the molecular mass fraction is approximately given by
\begin{equation}
\label{fh2approx}
f_{\rm H_2}(N_{\rm c}, Z') \approx 1 - \left[1 + \left(\frac{3}{4} \frac{s}{1+\delta}\right)^{-5}\right]^{-1/5},
\end{equation}
where $N_{\rm c}$ is the column density of cold gas (i.e.\ including CNM and molecular gas, but excluding warm atomic gas), $s = \ln(1 + 0.6\chi)/(0.045 N_{20} Z')$, $N_{20} = N_{\rm c}/[10^{20}\mbox{ H nuclei cm}^{-2}]$  and $\delta=0.0712(0.1 s^{-1}+0.675)^{-2.8}$. (This approximation is slightly different than that given in \citetalias{krumholz09a}; the two agree closely for clouds that are substantially molecular, but this one is more accurate at low $f_{\rm H_2}$ -- McKee \& Krumholz, 2009, in preparation.) This expression gives the mass of the spherical molecular core of the cloud, which is surrounded by a shell of atomic gas whose density is lower than that of the molecular gas by a factor $\phi_{\rm mol}$, which \citetalias{krumholz09a} show is typically $\simeq 10$. The covering fraction of the molecular sphere is
\begin{equation}
\label{coveringfrac}
c_{\rm H_2} = \left[1 - \phi_{\rm mol} \left(1-f_{\rm H_2}^{-1}\right)\right]^{-2/3}.
\end{equation}
Obviously a spherical ball is a great oversimplification of the complex geometries of atomic-molecular complexes, but $c_{\rm H_2}$ is a useful general indicator of the fraction of the area that is likely to be covered by molecular material.

For a given metallicity $Z'$ it is trivial to numerically invert equation (\ref{coveringfrac}) to calculate the total cold gas column density $N_{\rm c}$ for which the molecular covering fraction reaches a particular value $c_{\rm H_2}$. The mean atomic column density is then $N(\hi) = (1-f_{\rm H_2}) N_{\rm c}$. This defines a locus of points in the $N(\hi), Z'$-plane corresponding to the specified $c_{\rm H_2}$. The maximum \hi\ column density corresponds to the limit $c_{\rm H_2} \rightarrow 1$, because this corresponds to an infinite slab illuminated by the external radiation field. At this point we must mention two important caveats. One is that we assume that the atomic gas in DLAs is in two-phase equilibrium, which may not be true for all of them. The other is that this method allows us to constrain only the cold \hi\ column density. In principle much larger warm gas column densities are possible (\S~\ref{twophase}).

\subsection{Comparison of Models and Observations}
\label{obscomp}

\begin{figure}
\plotone{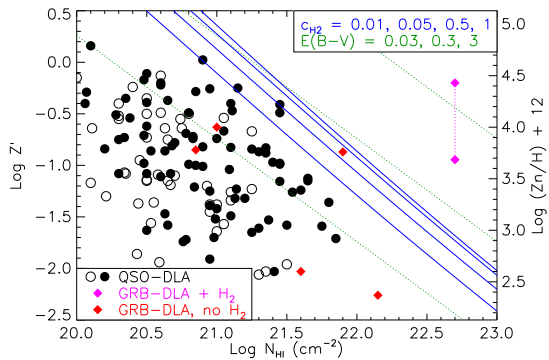}
\caption{
\label{dlamol_nhz}
\hi\ column density $N(\hi)$ versus normalized metallicity $Z'$ and zinc abundance $\log(\mbox{Zn}/\mbox{H})+12$, computed for cold gas. We show lines of constant H$_2$ covering fraction ({\it solid blue}, $c_{\rm H_2}$ increasing with $N(\hi)$), lines of constant color excess ({\it dotted green}, $E(B-V)$ increasing with $N(\hi)$), QSO-DLAs from \citet{herbert-fort06a}, Kaplan et~al.\ (2009, in preparation), and Dessauges-Zavadsky et~al.\ (2009, in preparation) ({\it black circles}), GRB-DLAs without H$_2$ detections ({\it red diamonds}; \citealt{prochaska07a}), and the GRB080607-DLA with an H$_2$ detection ({\it purple diamonds with line}, \citealt{prochaska09a}). For the QSO-DLAs, filled circles indicate detections of metals and open circles indicate the $1\sigma$ upper limits on metallicity. See the main text for discussion.}
\end{figure}

In Figure \ref{dlamol_nhz}, we plot our derived values $N(\hi)$ versus $Z'$ for molecular covering fractions from $c_{\rm H_2}=0.01-1$.\footnote{For $c_{\rm H_2}=1$ we use the KMT formalism to solve for $f_{\rm H_2}$ numerically rather than using equation (\ref{fh2approx}). This is necessary because as $c_{\rm H_2}\rightarrow 1$, $N(\hi)\rightarrow (1-f_{\rm H_2})N_{\rm c}$ depends on depends on $df_{\rm H_2}/dN_{\rm c}$. Equation (\ref{fh2approx}) does not give precisely the correct limit for this quantity.} We also show lines of constant $E(B-V)$, computed using a \citet{draine03a} $R_V=3.1$ extinction curve scaled by metallicity, giving $E(B-V)/N(\hi) = 1.65\times 10^{-22} Z'$ cm$^2$ and $A_V/N(\hi)=5.32\times 10^{-22} Z'$ mag cm$^2$. We compare to observed QSO- and GRB-DLAs from \citet{herbert-fort06a}, \citet{prochaska07a, prochaska09a}, Kaplan et~al.\ (2009, in preparation), and Dessauges-Zavadsky et~al.\ (2009, in preparation). For the Dessauges-Zavadsky et~al.\ sample we derive metallicities from zinc abundance: $\log Z' = [{\rm Zn}/{\rm H}] \equiv \log({\rm Zn}/{\rm H}) - \log ({\rm Zn}/{\rm H})_{\odot}$, where $\log({\rm Zn}/{\rm H})_{\odot} + 12 = 4.63$ \citep{lodders03a}. For all other data we use the metallicity reported by the authors. To avoid possible issues arising from either ionization correction or metallicity evolution with redshift, we exclude DLAs with $\log N(\hi)<20$ and redshift $z<1.7$.

As the Figure shows, the zone at high $N(\hi)$ and $Z'$ where no DLAs lie (except that associated with GRB080607, which we discuss below), corresponds well to the predicted zone of exclusion.  The molecular covering fraction declines sharply away from the $c_{\rm H_2} = 1$ line, so all DLAs but GRB080607 lie below $c_{\rm H_2}=0.06$. This is consistent with the results of \citet{zwaan06a}, who conclude that detection of true molecular clouds in DLAs is unlikely because the molecular material has a small covering fraction. Trace amounts of molecular hydrogen have been discovered in some DLAs \citep[e.g.][]{ledoux03a, noterdaeme08a}, but these low molecular columns almost certainly correspond to H$_2$ spatially mixed with cold atomic gas, rather than true molecular clouds. The KMT formalism approximates the atomic-molecular transition as sharp, so it does not apply to these systems. We defer discussion of them to future work. Also note that the observed distribution falls off sharply at $\log Z' \ga 0$, and at $\log N(\hi) \ga 22$ independent of $Z'$. Molecule formation cannot explain these features.

\subsection{The DLA Associated with GRB080607}
\label{grbdla}

\begin{figure}
\plotone{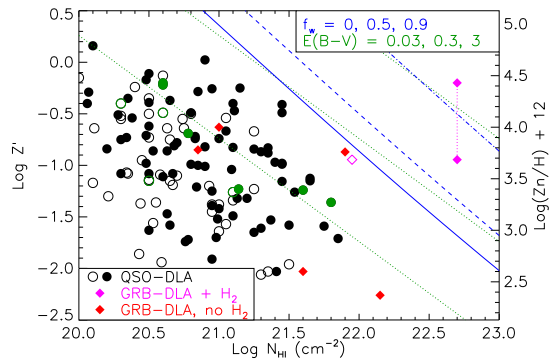}
\caption{
\label{dlamol_warm}
Same as Figure \ref{dlamol_nhz}, plus blue lines showing the maximum values of $N(\hi)$ and $Z'$ assuming that warm gas fractions $f_{\rm w}=0.0$ ({\it solid}), $f_{\rm w}=0.5$ ({\it dashed}), and $f_{\rm w}=0.9$ ({\it dot-dashed}). QSO-DLA points shown in green are those with measured spin temperatures \citep{kanekar09a}. The open purple diamond shows the estimated GRB080607-DLA cold \hi\ column (see \S~\ref{grbdla}).
}
\end{figure}

The DLA associated with GRB080607 \citep{prochaska09a}, the only DLA inside the zone of exclusion, is also the only DLA to show significant columns of H$_2$ and CO. We plot this detection at two metallicities derived in different ways. The $\log Z'=-0.2$ point corresponds to the oxygen abundance $[{\rm O}/{\rm H}]$. The $\log Z'=-0.94$ point is derived using the observed visual extinction $A_V\approx 3.2$ and \hi\ and H$_2$ columns $\log N(\hi)=22.7$ and $\log N({\rm H}_2)=21.2$, and adopting the same metallicity-dependent $A_V/N(\hi)$ ratio as in \S~\ref{obscomp}. Since the molecular fraction depends on solids that can catalyze H$_2$ formation and absorb LW photons, the latter estimate is probably the relevant one. The solid content of the atomic gas may be even lower if a disproportionate share of the observed extinction comes from the molecular material.

While the detection of molecules is consistent with the DLA's presence in the zone of exclusion, we have not yet explained why its column is only $6\%$ molecular. Dissociation by the GRB afterglow is unlikely to be the explanation. The hard afterglow spectrum would produce nearly coincident ionization and dissociation fronts with little atomic hydrogen between them \citep{draine02a}, and excited H$_2$ just outside the dissociation front would produce strong absorption features in the DLA that are not observed (\citealt{prochaska09a}; however, see Sheffer et~al., 2009, in preparation). Observations instead suggest that the molecular cloud is at least 100 pc from the GRB, so our line of sight must pass through the host galaxy's disk at a glancing angle. Thus a majority of the material along the line of sight should be warm atomic gas that is unrelated to the molecular cloud, and, as discussed in \S~\ref{theory}, provides no shielding to it.

Let the DLA sightline contain warm  and cold (atomic plus molecular) material with column densities $N_{\rm w}$ and $N_{\rm c}$. The molecular column will be
\begin{equation}
1.8 N({\rm H}_2) = f_{\rm H_2}(N_{\rm c}, Z') N_{\rm c},
\end{equation}
where the factor of $1.8$ accounts for the difference in mean number of particles per unit mass between atomic and molecular gas. For $\log Z'=-0.94$ and $\log N({\rm H}_2) = 21.2$, solving this equation yields $\log N_{\rm c} = 22.0$, which implies $f_{\rm H_2} = 1.8 N_{\rm H_2}/N_{\rm c}=0.31$ and $\log N_{\rm w}=22.6$. Thus we can explain the observed \hi\ and H$_2$ column densities if 18\% of the \hi\ column consists of cold gas and the remaining 82\% is warm. The open purple diamond in Figure \ref{dlamol_warm} shows where this system would fall in the $N(\hi),Z'$ plane if we counted only cold \hi. The path length through the warm gas is $L= 13 (n_{\rm WNM}/1\mbox{ cm}^{-3})^{-1}$ kpc, where $n_{\rm WNM}$ is the volume density. For $n_{\rm WNM}\sim 10^{-0.5}$ cm$^{-3}$, a typical value in the Solar neighborhood, this would require $L \sim 40$ kpc, but in a two-phase medium the equilibrium WNM density is close to linearly proportional to the intensity of the FUV radiation field in the galaxy \citep{wolfire03}. Since the FUV radiation field in other GRB host galaxies is $\sim 10-100$ times larger than the Solar neighborhood value \citep{tumlinson07a}, we expect $n_{\rm WNM}\sim 10$ cm$^{-3}$, giving $L\sim 1$ kpc.

\section{Constraining the Warm Gas Fraction}
\label{twophase}

Warm gas adds \hi\ column without creating molecules, possibly moving systems to the right into the exclusion zone. To illustrate this, in Figure \ref{dlamol_warm} we show the same observed systems as in Figure \ref{dlamol_nhz}, compared with the line $c_{\rm H_2}=1$ calculated assuming that a fraction $f_{\rm w}$ of the observed \hi\ column is in the form of warm gas that provides no shielding. These lines indicate the maximum values of $N(\hi)$ that can be observed for a given metallicity $Z'$ for the indicated value of $f_{\rm w}$. 

DLAs with large total gas columns, high $Z'$, and large $f_{\rm w}$, would not form molecules, so they would be observed to the right of the $f_{\rm w}=0$ line. The fact that this region is unpopulated by QSO-DLAs suggests that the combination of large gas column, large $Z'$ and large $f_{\rm w}$ must be very rare for them. It may be more common for GRB-DLAs, since one of six systems falls within the exclusion zone. Of course even if such large column-large $Z'$ DLAs did exist, the distribution would likely be cut off at some point by dust extinction effects, and even in the absence of CNM a sufficiently large WNM column could provide enough shielding for molecules to form. A precise estimate of the warm gas fraction permitted by the data would require a careful analysis of these effects. Nonetheless, the fact that dust extinction alone cannot explain the zone of exclusion strongly argues that $f_{\rm w}$ cannot be anywhere near unity for the DLAs with the highest values of $N(\hi)$ and $Z'$. Our conclusion is consistent with that of \citet{wolfe08a}, who argue based on C~\textsc{ii}$^*$ absorption measurements that $\sim 1/2$ of DLAs contain significant cold gas. 

However, we emphasize that our model offers no constraints on the warm gas fraction in DLAs with total column densities and metallicities that place them below the $f_{\rm w} = 1$ line. Thus our results are consistent with the conclusion of \citet{kanekar09a}, who find large values of $f_{\rm w}$ for the DLAs shown by the green points in Figure \ref{dlamol_warm}. Significantly, all of these points are at least $1.6$ dex to the left of the $f_{\rm w}=0$ line.


\section{Summary and Discussion}
\label{disc}

We show that the absence of high column density, high metallicity DLAs results from the conversion of atomic into molecular gas, confirming the proposal of \citet{schaye01a}. Our model is insensitive to the DLA radiation environment due to the way that two-phase atomic gas responds to variations in radiation field. The maximum metallicity and column density do depend on the fraction of the atomic gas that is cold and dense, and the observed distribution of QSO-DLAs is inconsistent with the existence of a significant population of high column density, high metallicity QSO-DLAs dominated by warm gas.

Given our conclusion that high column density DLAs must host significant amounts of cold gas, one might ask why, as pointed out by \citet{wolfe06a}, most DLAs cannot host significant {\it in situ} star formation. The answer is that the presence of cold gas is a necessary but not sufficient condition for star formation. Although DLAs populate the $N(\hi), Z'$ plane up to the point where the molecule fraction becomes large, the vast majority of them are found at much lower column densities and metallicities, where they are not expected to have any significant amount of molecular gas. If stars form exclusively in molecular gas, as numerous observations now seem to suggest, then the vast majority of DLA columns should be inert as far as star formation is concerned.

This is not to say that DLA systems do not host any star formation. Indeed, indirect measures of the radiation fields in some QSO-DLAs show strong evidence for the presence of a local heat source that is likely to be star formation \citep{wolfe08a}, and in GRB-DLA systems there is obviously evidence for ongoing star formation. Our result simply suggests that the star formation must be taking place in other parts of these galaxies, which have significantly higher column densities and molecular contents than the sightlines we most commonly observe as DLAs.

\acknowledgements We thank C.~F.\ McKee and the referee, J.\ Schaye, for helpful comments. Support for this work was provided by the Alfred P.\ Sloan Foundation (MRK), by NASA/JPL through the Spitzer Theoretical Research Program (MRK), by the National Science Foundation through grants AST-0807739 (MRK) and AST-0709235 (JXP), and by an NSERC Discovery Grant (SLE).


\end{document}